\documentclass[12pt]{article}
\usepackage{amsmath}
\usepackage{amssymb,amsthm}
\usepackage{graphicx,epsfig}
\usepackage[english]{babel}
\usepackage[textwidth=18.5cm,textheight=22cm]{geometry}
\newcommand{\C}{\mathbb{C}}

\newcommand{\Z}{\mathbb{Z}}

\newcommand{\mS}{\mathbb{S}}

\renewcommand{\d}{\mathrm{d}}
\newcommand{\ct}{\boldsymbol{\mathrm{t}}}
\newcommand{\dt}{\bar{\boldsymbol{\mathrm{t}}}}
\newcommand{\bt}{\boldsymbol{t}}

\newcommand{\be}{\begin{equation}}
\newcommand{\ee}{\end{equation}}

\newtheorem{teh}{Theorem}

\begin{document}

\title{ A common integrable structure in the  hermitian matrix model
and Hele-Shaw flows
\thanks{Partially supported by  MEC
project FIS2005-00319 and ESF programme MISGAM}}
\author{L. Mart\'{\i}nez Alonso$^{1}$ and E. Medina$^{2}$
\\
\emph{$^1$ Departamento de F\'{\i}sica Te\'{o}rica II, Universidad
Complutense}\\ \emph{E28040 Madrid, Spain}\\
\emph{$^2$ Departamento de Matem\'aticas, Universidad de C\'adiz}\\
\emph{E11510 Puerto Real, C\'adiz, Spain} }
\date{} \maketitle
\maketitle \abstract{It is proved that the system of string equations of the
dispersionless 2-Toda hierarchy which arises in the planar limit of
the hermitian matrix model also underlies certain processes in
Hele-Shaw flows.

}

\section{Introduction}
The Toda hierarchy represents a relevant integrable structure which
emerges in several random matrix models \cite{ger}-\cite{avm}. Thus,
the partition functions
\begin{equation}\label{1}
Z_N(\mbox{Hermitian})=\int \d H \exp\Big(\mbox{tr}(\sum_{k\geq
1}t_k\,H^k)\Big),
\end{equation}
\begin{equation}\label{2}
Z_N(\mbox{Normal})=\int \d M\,\d M^{\dagger}
\exp\Big(\mbox{tr}\,(M\,M^{\dagger}+\sum_{k\geq
1}(t_k\,M^k+\bar{t}_k\,M^{\dagger\,k}))\Big),
\end{equation}
of the hermitian ($H=H^{\dagger}$) and the normal matrix models
($[M,M^{\dagger}]=0$) , where $N$ is the matrix dimension,  are
tau-functions of the 1-Toda and 2-Toda hierarchy, respectively. As a
consequence of this connection  new facets of the Toda hierarchy
have been discovered. Thus the analysis of the large $N$-limit of
the Hermitian matrix model lead to introduce an interpolated
continuous version of the 2-Toda hierarchy: the \emph{dispersionful}
2-Toda hierarchy (see for instance \cite{tt}). On the other hand,
the leading contribution to the large $N$-limit (planar
contribution) motivated the introduction of a \emph{classical}
version of the Toda hierarchy \cite{tt} which is known as the
\emph{dispersionless} 2-Toda (d2-Toda) hierarchy.

Laplacian growth processes describe evolutions of two-dimensional
domains driven by harmonic fields. It was shown in \cite{zab1} that
the d2-Toda is a relevant integrable structure in Laplacian growth
problems and conformal maps dynamics. For example, if  a given
analytic curve $\gamma\, \,(z=z(p),\, |p|=1)$ is the boundary of a
simply-connected bounded domain,  then $\gamma$ evolves with respect
to its harmonic moments according to a solution of the d2-Toda
hierarchy. These  solutions are characterized by the string
equations
\begin{equation}\label{11}
\bar{z}=m,\quad \overline{m}=-z.
\end{equation}
Here $(z,m)$ and $(\bar{z},\bar{m})$ denote the two pairs of
Lax-Orlov operators of the d2-Toda hierarchy. As it was noticed in
\cite{zab1}-\cite{wz}, this integrable structure also emerges in the
planar limit of the normal matrix model \eqref{2} and describes the
evolution of the support of eigenvalues under a change of the
parameters $t_k$ of the potential.

The present paper is motivated by the recent discovery \cite{lee} of
an integrable structure provided by the dispersionless AKNS
hierarchy which describes the bubble break-off in Hele-Shaw flows.
In this work we prove that this integrable structure is also
characterized  by the solution of a pair of string equations
\begin{equation}\label{trii}
z=\bar{z},\quad m=\overline{m},
\end{equation}
of the d2-Toda hierarchy. Since the system \eqref{trii} describes
the planar limit of \eqref{1}, it constitutes a common integrable
structure arising in the Hermitian matrix model and the theory of
Hele-Shaw flows.

 Our strategy is inspired by previous results \cite{mel3}-\cite{mano2} on solution
methods for dispersionless string equations. We also develop some
useful standard technology of the theory of Lax equations in the
context of the d2-Toda hierarchy.

The paper is organized as follows:

In the next section  the basic theory of the d2-Toda hierarchy, the
method of  string equations and the solution of \eqref{trii} are
discussed. In Section 3 we show how the solution of \eqref{trii}
appears in the planar limit of the Hermitian matrix model and the
Hele-Shaw bubble break-off processes studied in \cite{lee}.

\section{The dispersionless Toda
hierarchy }

\subsection{String equations in the d2-Toda hierarchy}

The dispersionless d2-Toda hierarchy\cite{tt} can be formulated in
terms of two pairs $(z,m)$ and $(\bar{z},\overline{m})$ of Lax-Orlov
functions, where $z$ and $\bar{z}$  are series in a  complex
variable $p$ of the form
\begin{equation}\label{d0a}
z=p+u+\dfrac{u_1}{p}+\cdots,\quad
\bar{z}=\dfrac{v}{p}+v_0+v_1\,p+\cdots,
\end{equation}
while $m$ and $\bar{m}$ are series in $z$ and $\bar{z}$ of the form
\begin{equation}\label{act0}
m:=\sum_{j=1}^\infty j\, t_j z^{j-1}+\dfrac{x}{z}+\sum_{j\geq
1}\dfrac{S_{j+1}}{z^{j+1}} ,\quad \overline{m} :=\sum_{j=1}^\infty
j\,\bar{t}_j \bar{z}^{j-1}-\dfrac{x}{\bar{z}}+\sum_{j\geq
1}\dfrac{\bar{S}_{j+1}}{\bar{z}^{j+1}}.
\end{equation}
The coefficients in the expansions \eqref{d0a} and \eqref{act0}
depend on a complex variable $x$ and two infinite sets of complex
variables $\ct:=(t_1,t_2,\ldots)$ and
$\dt:=(\bar{t}_1,\bar{t}_2,\ldots)$. The d2-Toda hierarchy is
encoded in the equation
\begin{equation}\label{2.a}
\d z\wedge\d m=\d \bar{z}\wedge\d \overline{m}= \d \Big( \log{p}\,\d
x+ \sum_{j=1}^\infty\Big( (z^{j})_{+}\,\d t_j+(\bar{z}^{j})_{-}\,\d
\bar{t}_j\Big)\Big).
\end{equation}
Here the $(\pm)$ parts  of $p$-series denote  the truncations in the
positive and strictly negative power terms, respectively. As a
consequence there exist two \emph{action} functions $S$ and
$\bar{S}$ verifying
\begin{align*}
 \d S&=m\,\d z+\log{p}\,\d x+ \sum_{j=1}^\infty\Big(
(z^{j})_{+}\,\d t_j+(\bar{z}^{j})_{-}\,\d \bar{t}_j\Big), \\
 \d\bar{S}&=\overline{m}\,\d \bar{z}+\log{p}\,\d
x+\sum_{j=1}^\infty \Big( (z^{j})_{+}\,\d t_j+(\bar{z}^{j})_{-}\,\d
\bar{t}_j\Big),
\end{align*}
and such that they admit expansions
\begin{equation}\label{act}
S =\sum_{j=1}^\infty t_j z^{j}+x\,\log{z}-\sum_{j\geq
1}\dfrac{S_{j+1}}{jz^{j}},\quad \bar{S} =\sum_{j=1}^\infty \bar{t}_j
\bar{z}^{j}-x\,\log{\bar{z}}-\bar{S}_0+\sum_{j\geq
1}\dfrac{\bar{S}_{j+1}}{j\bar{z}^{j+1}}.
\end{equation}
From \eqref{2.a} one derives  the d2-Toda  hierarchy in Lax form
\begin{equation}\label{d3}
\dfrac{\partial \mathcal{K}}{\partial
t_j}=\{(z^j)_+,\mathcal{K}\},\quad \dfrac{\partial
\mathcal{K}}{\partial \bar{t}_j}=\{(\bar{z}^j)_-,\mathcal{K}\},
\end{equation}
where $\mathcal{K}=z,\,m,\,\bar{z},\,\overline{m}$,  and we are
using the Poisson bracket $\{f,g\}:=p\,(f_p\,g_x-f_x\,g_p)$.

\vspace{0.3cm}

 The
following result was proved by Takasaki and Takebe (see \cite{tt}):

\begin{teh}

Let $(P(z,m),Q(z,m))$ and
$(\overline{P}(\bar{z},\overline{m}),\overline{Q}
(\bar{z},\overline{m})))$ be functions such that
\[
\{P,Q\}=\{z,m\} ,\quad
\{\overline{P},\overline{Q}\}=\{\bar{z},\overline{m}\}.
\]
If $(z,m)$ and $(\bar{z},\overline{m})$ are functions which can be
expanded in the form \eqref{d0a}-\eqref{act0} and satisfy the pair
of constraints
\begin{equation}\label{dstring}
P(z,m)=\overline{P}(\bar{z},\overline{m}),\quad
Q(z,m)=\overline{Q}(\bar{z},\overline{m}) ,
\end{equation}
then they verify $\{z,m\}=\{\bar{z},\overline{m}\}=1$ and are
solutions of the Lax equations \eqref{d3} for the d2-Toda hierarchy
.

\end{teh}

Constraints of the form \eqref{dstring} are called
\emph{dispersionless string equations}. In this paper we are
concerned with the system \eqref{trii}. The first equation
$z=\bar{z}$ of \eqref{trii} defines the 1-Toda reduction of the
d2-Toda hierarchy
\begin{equation}\label{d4}
z=\bar{z}=p+u+\dfrac{v}{p},
\end{equation}
where
\begin{equation}\label{d6}
u=\partial_x S_2,\quad \log{v}=-\partial_x\bar{S}_0.
\end{equation}
As a consequence  the Lax equations \eqref{d3} imply that $u$ and
$v$ depend on $(\ct,\dt)$ through the combination $\ct-\dt$.

Due to \eqref{d4} there are two branches of $p$ as a function of $z$
\begin{align}\label{d5}
\nonumber
&p(z)=\dfrac{1}{2}\Big((z-u)+\sqrt{(z-u)^2-4v}\Big)=z-u-\dfrac{v}{z}+\cdots\\\\
\nonumber
&\bar{p}(z)=\dfrac{1}{2}\Big((z-u)-\sqrt{(z-u)^2-4v}\Big)=\dfrac{v}{z}+\cdots.
\end{align}
To characterize the members of the d1-Toda hierarchy of integrable
systems as well as to solve the string equations \eqref{trii} it is
required to determine $(z^j)_{-}(p(z))$ and $(z^j)_{+}(\bar{p}(z))$
in terms of $(u,v)$. By using \eqref{d4} it is clear that there are
functions $\alpha_j,\,\beta_j,\,\bar{\alpha}_j, \, \bar{\beta}_j)$,
which depend polynomially, in $z$   such that
\begin{align*}
\partial_{t_j} S(z)=(z^j)_+(p(z))&=\alpha_j+\beta_j\,p(z),\quad
\partial_{\bar{t}_j} S(z)=(z^j)_-(p(z))=\bar{\alpha}_j+\bar{\beta}_j\,p(z),\\
\partial_{t_j} \bar{S}(z)=(z^j)_+(\bar{p}(z))&=\alpha_j+\beta_j\,\bar{p}(z),\quad
\partial_{\bar{t}_j} \bar{S}(z)=(z^j)_-(\bar{p}(z))=\bar{\alpha}_j+\bar{\beta}_j\,\bar{p}(z),
\end{align*}
and
\begin{equation}\label{d8}
\bar{\alpha}_j=z^j-\alpha_j,\quad \bar{\beta}_j=-\beta_j.
\end{equation}
Now we have
\begin{equation}\label{d7}
\alpha_j+\beta_j\,p(z)=\partial_{t_j} S(z)=
z^j+\mathcal{O}\Big(\dfrac{1}{z}\Big),\quad
\alpha_j+\beta_j\,\bar{p}(z)=\partial_{t_j}\bar{S}(z)
=-\partial_{t_j} \bar{S}_0+ \mathcal{O}\Big(\dfrac{1}{z}\Big),
\end{equation}
so that
\begin{equation}\label{d7aa}
\alpha_j=\dfrac{1}{2}\Big(z^j-\partial_{t_j}
\bar{S}_0-(p+\bar{p})\,\beta_j\Big),\quad
\beta_j=\Big(\dfrac{z^j}{p-\bar{p}}\Big)_\oplus,
\end{equation}
where $(\;)_\oplus$ and $(\;)_\ominus$  stand for the projection of
$z$-series on the positive and strictly negative powers,
respectively. Thus, by introducing the generating function
\begin{equation}\label{d7b}
R :=\dfrac{z}{p-\bar{p}}=\dfrac{z}{\sqrt{(z-u)^2-4v}}=\sum_{k\geq
0}\dfrac{r_k(u,v)}{z^k},\quad r_0=1.
\end{equation}
we deduce
\begin{align}\label{d7c}
\nonumber
(z^j)_+(p(z))&=z^j-\dfrac{1}{2}\,\partial_{t_j} \bar{S}_0-\dfrac{z}{2\,R}\,\Big(z^{j-1}\,R\Big)_\ominus\\
&=z^j-\dfrac{1}{2}(\partial_{t_j}
\bar{S}_0+r_j)-\dfrac{1}{2\,z}\,(r_{j+1}-u\,r_j)+\mathcal{O}\Big(\dfrac{1}{z^2}\Big).
\end{align}
Hence
\[
\partial_{t_j} \bar{S}_0=-r_j,\quad \partial_{t_j}
S_2=\dfrac{1}{2}\,(r_{j+1}-u\,r_j),
\]
so that the equations of the $d1$-Toda hierarchy are given by
\begin{equation}\label{dto1}
\partial_{t_j} u=\dfrac{1}{2}\,\partial_x\,(r_{j+1}-u\,r_j),\quad
\partial_{t_j} v=v\,\partial_x\,r_{j}.
\end{equation}

Furthermore, we have found
\begin{equation}\label{d7d0}
(z^j)_-(p(z))=
-\dfrac{1}{2}\,r_j+\dfrac{z}{2\,R}\,\Big(z^{j-1}\,R\Big)_\ominus,
\quad (z^j)_+(\bar{p}(z))=r_j+(z^j)_-(p(z)).
\end{equation}
Hence, the first terms of their asymptotic expansions as
$z\rightarrow\infty$ are
\begin{equation}\label{d7d}
(z^j)_-(p(z))=\dfrac{1}{2\,z}\,(r_{j+1}-u\,r_j)+\mathcal{O}\Big(\dfrac{1}{z^2}\Big),
\quad
(z^j)_+(\bar{p}(z))=r_j+\dfrac{1}{2\,z}\,(r_{j+1}-u\,r_j)+\mathcal{O}\Big(\dfrac{1}{z^2}\Big).
\end{equation}
Notice that since $r_0=1$ and $ r_1=u$, these last equations hold
for $j\geq 0$.

\subsection{Hodograph solutions of the $1$-dToda hierarchy}

In the above paragraph we have used  the first string equation of
\eqref{trii}. Let us now deal with the second one. To this end we
set
\[
m=\overline{m}= \sum_{j=1}^\infty j\,t_j\,(z^{j-1})_++
\sum_{j=1}^\infty j\,\bar{t}_j\,(z^{j-1})_-,
\]
which leads to the following expressions for the Orlov functions
$(m,\overline{m})$
\begin{align}\label{d8a}
\nonumber &m(z)=\sum_{j=1}^\infty j\,t_j\,z^{j-1}+\sum_{j=1}^\infty
j\,(\bar{t}_j-t_j)\,(z^{j-1})_-(p(z)),\\\\
\nonumber &\overline{m}(z)=\sum_{j=1}^\infty
j\,\bar{t}_j\,z^{j-1}-\sum_{j=1}^\infty
j\,(\bar{t}_j-t_j)\,(z^{j-1})_+(\bar{p}(z)).
\end{align}
In order to apply Theorem 1 we have to determine $u$ and $v$ and
ensure that $(m,\overline{m})$ verify the correct asymptotic form
\eqref{d0a}-\eqref{act0}. Both things can be achieved  by reducing
\eqref{d8a} to the form
\begin{align}\label{d9}
\nonumber &\dfrac{x}{z}+\sum_{j\geq
2}\dfrac{1}{z^j}S_j=\sum_{j=1}^\infty
j\,(\bar{t}_j-t_j)\,(z^{j-1})_-(p(z)),\\\\
\nonumber &-\dfrac{x}{z}+\sum_{j\geq
2}\dfrac{1}{z^j}\bar{S}_j=-\sum_{j=1}^\infty
j\,(\bar{t}_j-t_j)\,(z^{j-1})_+(\bar{p}(z)),
\end{align}
and equating coefficients of powers of $z$.  Indeed, from
\eqref{d7d} we see that identifying the coefficients of $z^{-1}$ in
both sides of the two equations of \eqref{d9} yields the same
relation. This equation together with the one supplied by
identifying the coefficients of the constant terms in the second
equation of \eqref{d9} provides the following system of
\emph{hodograph-type} equations to determine $(u,v)$
\begin{equation}\label{ho}\begin{cases}
\sum_{j=1}^\infty j\,(\bar{t}_j-t_j)
r_{j-1}=0,\\\\
\dfrac{1}{2}\,\sum_{j=1}^\infty j\,(\bar{t}_j-t_j)\Big)\,r_j=x.
\end{cases}
\end{equation}
It can be rewritten as
\begin{equation}\label{hoin}\everymath{\displaystyle}
\begin{cases}
\oint_{\gamma}\dfrac{dz}{2\pi i}
\dfrac{V_{z}}{\sqrt{(z-u)^2-4v}}\, =0,\\\\
\oint_{\gamma}\dfrac{dz}{2\pi
i}\dfrac{z\,V_{z}}{\sqrt{(z-u)^2-4v}}\, =-2\,x,
\end{cases}
\end{equation}
where  $\gamma$ is a large enough positively oriented closed path
and $V_{z}$ denotes the derivative with respect to $z$ of the
function
\begin{equation}\label{U}
V(z,\ct-\dt):=\sum_{j=1}^\infty (t_j-\bar{t}_j)\Big)\,z^j.
\end{equation}
The remaining equations arising from \eqref{d9} characterize the
functions $S_j^{(0)}$ and $\overline{S}_j^{(0)}$ for $j\geq 1$ in
terms of $(u,v)$. Therefore we have characterized a solution $(z,m)$
and $(\bar{z},\overline{m})$ of the system of string equations
\eqref{trii} verifying the conditions of Theorem 1 and,
consequently, it solves the d1-Toda hierarchy.

\section{Planar limit of the Hermitian matrix model and bubble break-off in Hele-Shaw flows}

\subsection{The Hermitian matrix model}

 If we write the partition function \eqref{1} of the Hermitian matrix model in
terms of eigenvalues and slow variables $\ct:=\epsilon\,\bt$, where
$\epsilon=1/N$, we get
\begin{equation}\label{mat}
Z_n(N\,\ct)=\int_{\mathbb{R}^n}\prod_{k=1}^{n}\Big(d\,x_k\,e^{N\,V(x_k,\ct)})\Big)(\Delta(x_1,\cdots,x_n))^2,\quad
V(z,\ct):=\sum_{k\geq 1}t_k\,z^k.
\end{equation}
The  large $N$-limit of the model is determined by the asymptotic
expansion  of $Z_n(N\,\ct) $ for $n=N$ as $N\rightarrow \infty$
\begin{equation}
Z_N(N\,\ct)
=\int_{\mathbb{R}^N}\prod_{k=1}^{N}\Big(d\,x_k\,e^{N\,V(x_k,\ct)})\Big)(\Delta(x_1,\cdots,x_N))^2,
\end{equation}
 It is well-known \cite{avm} that   $Z_n(\bt)$ is a $\tau$-function of the semi-infinite 1-Toda hierarchy , then
there exists a $\tau$-function $\tau(\epsilon,x,\ct)$ of the
dispersionful 1-Toda hierarchy verifying
\begin{equation}\label{rel}
\tau(\epsilon,\epsilon\,n,\ct)=Z_n(N\,\ct),
\end{equation}
and consequently
\begin{equation}\label{rel1}
\tau(\epsilon,1,\ct)=Z_N(N\,\ct).
\end{equation}
Hence the large $N$-limit expansion of the partition function
\begin{equation}\label{tau1}
\Z_N(N\,\ct)=\exp{\Big(N^2\,\mathbb{F}\Big)},\quad
\mathbb{F}=\sum_{k\geq 0}\dfrac{1}{N^{2k}}\,F^{(2k)},
\end{equation}
is determined by  a solution of the dispersionful 1-Toda hierarchy
at $x=1$.

As a consequence of the above analysis one concludes that the
leading term (planar limit) $F^{(0)}$ is determined by a solution of
the 1-dToda hierarchy at $x=1$. Furthermore, the leading terms of
the $N$-expansions of the main objects of the hermitian matrix model
can be expressed in terms of quantities  of the  1-dToda hierarchy.
For example, in the \emph{one-cut} case , the density of eigenvalues
\[
\rho(z)=M(z)\,\sqrt{(z-a)(z-b)},
\]
is supported on a single interval $[a,b]$. These objects are related
to the leading term $W^{(0)}$ of the one-point correlator \cite{gin}
\[
W(z):=\dfrac{1}{N}\,\sum_{j\geq 0}\dfrac{1}{z^{j+1}}\langle tr
M^j\rangle=\dfrac{1}{z}+\dfrac{1}{N^2}\,\sum_{j\geq
1}\dfrac{1}{z^{j+1}}\, \dfrac{\partial \log\,Z_N(N\,\ct)}{\partial
t_j},
\]
in the form
\[
W^{(0)}=-\dfrac{1}{2}V_z(z)+i\pi\,\rho(z).
\]
On the other hand, it can be proved (see for instance \cite{eyn})
that
\begin{equation}
W^{(0)}=\label{m1} m(z,1,\ct)-\sum_{j=1}^\infty j\,t_j\,z^{j-1},
\end{equation}
so that \eqref{d7d0} and \eqref{d8a} yield
\begin{align}\label{e1}
\nonumber &-\dfrac{1}{2}V_z(z)+i\pi\,\rho(z)=-\sum_{j=1}^\infty
j\,t_j\,(z^{j-1})_-(p(z))\\\\
\nonumber &=\dfrac{1}{2}\sum_{j=1}^\infty
j\,t_j\,r_{j-1}-\dfrac{1}{2}\sum_{j=1}^\infty
j\,t_j\,z^{j-1}+\dfrac{1}{2}(p-\bar{p})\sum_{j=2}^\infty
j\,t_j\,\Big(z^{j-2}\,R\Big)_\oplus,
\end{align}
 Since we are setting
$\bar{t}_j=0,\,\forall j\geq 1$,  according to the first hodograph
equation \eqref{ho} the first term in the last equation vanishes.
Therefore the density of eigenvalues and  its support $[a,b]$ are
characterized by
\begin{align}\label{den}
\nonumber \rho(z)&:=\dfrac{1}{2\pi i}\Big(\dfrac{V_z}{\sqrt{(z-a)(z-b)}}\Big)_\oplus\,\sqrt{(z-a)(z-b)},\\\\
\nonumber a&:=u-2\,\sqrt{v},\quad b:=u+2\,\sqrt{v},
\end{align}
where we set $x=1$ in all the $x$-dependent functions. Observe that
according to \eqref{e1}
\begin{equation}\label{e2}
i\,\pi\,\rho(z)=\dfrac{1}{2}\,V_z(z)+\dfrac{x}{z}+\mathcal{O}\Big(\dfrac{1}{z^2}\Big),\quad
z\rightarrow\infty,
\end{equation}
so that the constraint $x=1$ means that the density of eigenvalues
is  normalized on its support
\[
\int_a^b\, \rho(z)\,dz=1.
\]

Moreover, from  \eqref{hoin} we obtain
\begin{equation}\label{ho3}
\oint_{\gamma}\dfrac{dz}{2\pi i}\dfrac{V_{z}}{\sqrt{(z-a)(z-b)}}\,
=0,\quad \oint_{\gamma}\dfrac{dz}{2\pi
i}\dfrac{z\,V_{z}}{\sqrt{(z-a)(z-b)}}\, =-2,
\end{equation}
with $\gamma$ being a positively oriented closed path encircling the
interval $[a,b]$. These are the quations which determine  the
zero-genus contribution or planar limit to the partition function of
the hermitian model \cite{eyn}-\cite{mig}.

\subsection{Bubble break-off in Hele-Shaw flows}

A Hele-Shaw cell is a narrow gap between two plates filled with two
fluids: say oil surrounding one or several bubbles of air. Let $D$
denote the domain in the complex plane $\C$ of the variable
$\lambda$ occupied by the air bubbles. By assuming that $D$ is an
\emph{algebraic domain} \cite{lee}, the boundary $\gamma$ of $D$ is
characterized by a \emph{Schwarz function} $\mS=\mS(\lambda)$ such
that
\begin{equation}\label{sch}
\lambda^*=\mS(\lambda),\quad \lambda\in\gamma.
\end{equation}
The geometry of the domain $\C-D$ is completely encoded in $\mS$ and
it can be conveniently described in terms of the \emph{Schottky
double} \cite{wz}: a Riemann surface $\mathcal{R}$ resulting from
gluing two copies $H_{\pm}$ of $\C-D$ trough $\gamma$, adding two points
at infinity $(\infty,\overline{\infty})$ and defining the complex
coordinates
\[
\begin{cases}
\lambda_+(\lambda)=\lambda,\quad \lambda\in H_+,\\
\lambda_-(\lambda)=\lambda^*,\quad \lambda\in H_-.
\end{cases}
\]
In particular  $\mS\,d\lambda$ can be extended to a unique
meromorphic differential $\omega$ on $\mathcal{R}$.

The evolution of $\gamma$ is governed by  D'Arcy law: the velocity
in the oil domain is proportional to the gradient of the pressure.
In the absence of surface tension,  pressure is continuous across
$\gamma$ and then if the bubbles are assumed to be kept at zero
pressure, we are lead to the Dirichlet boundary problem
\begin{equation}\label{dir}
\begin{cases}
\bigtriangleup \mathcal{P}=0,\quad \mbox{on $\C-D$},\\
\quad \mathcal{P}=0 \quad \mbox{on $\gamma$},\\
\quad \mathcal{P}\rightarrow -\log|z|,\quad z\rightarrow\infty.
\end{cases}
\end{equation}
If one assumes D'Arcy law in the form
$\vec{v}=-2\,\vec{\nabla}\mathcal{P}$, then by introducing the
function
\begin{equation}\label{cpo}
\Phi(\lambda):=\xi(\lambda)+i\,\mathcal{P}(\lambda),
\end{equation}
where $\xi$ and $\mathcal{P}$ are the \emph{stream function} and the
pressure, respectively,  D'Arcy law can be rewritten as
\begin{equation}\label{dar}
\partial_t\, \mS=2\,i\,\partial_{\lambda}\,\Phi,
\end{equation}
where $t$ stands for the time variable.

 In the set-up considered in \cite{lee} air is drawn out from two fixed points of a simply-connected air bubble making the bubble  breaks into two emergent bubbles with highly curved tips.  Before the break-off the interface oil-air remains free of cusp-like singularities and develops a smooth neck. As it is shown in \cite{wz}-\cite{lee}, the condition for bubbles to be  at equal pressure implies that the integral
\[
\Pi:=\dfrac{1}{2}\oint_{\beta}\,\omega,
\]
where  $\omega$  is the  meromorphic extension of $\mS\,d\lambda$ to
$\mathcal{R}$ and $\beta$ is a cycle connecting the bubbles, is a
constant of the motion. Since at break-off $\beta$ contracts to a
point, it is obvious that a necessary condition for break-off is
that $\Pi$ vanishes.

\begin{center}
\centerline{\epsfig{file=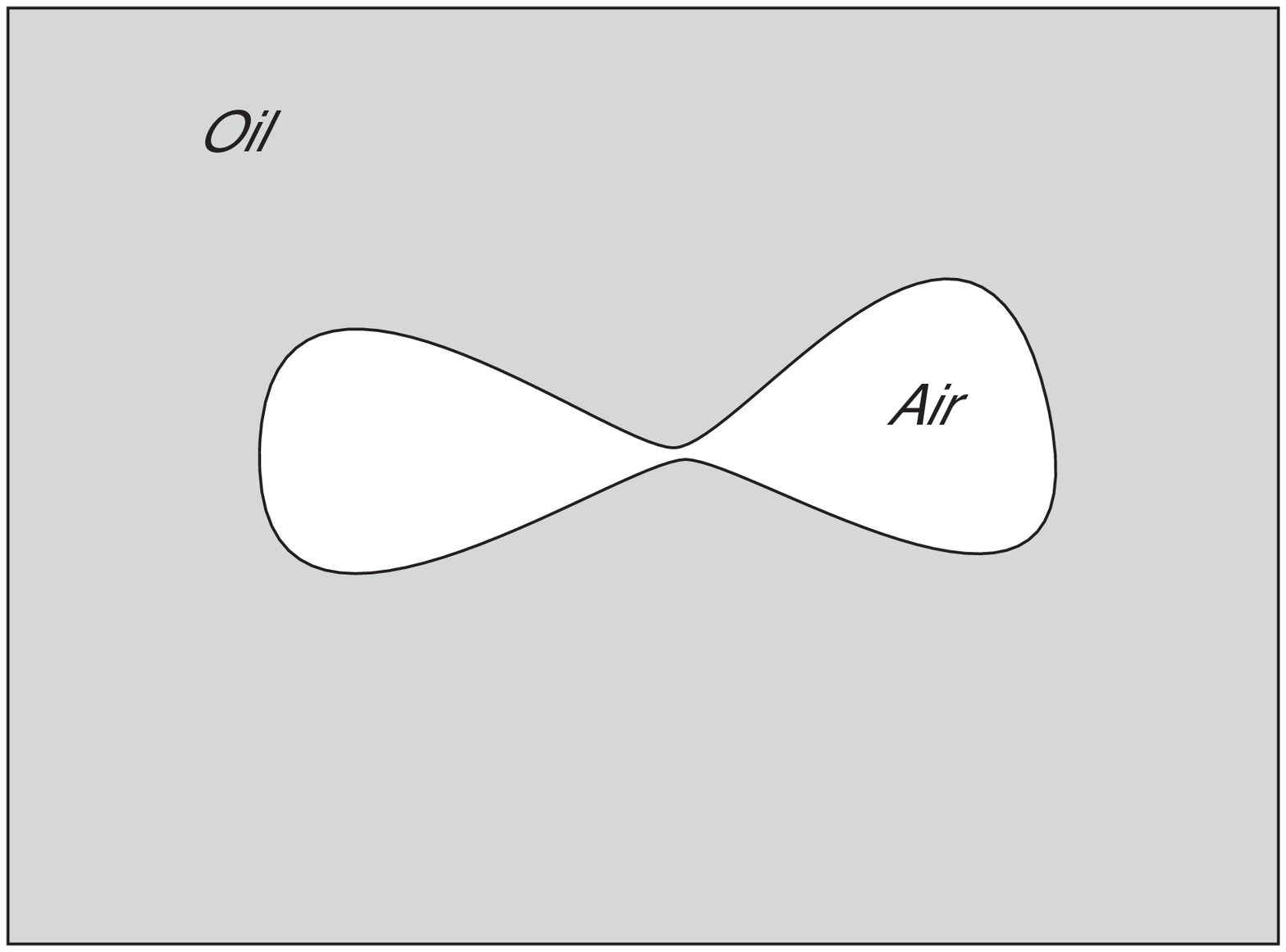,width=9cm}}
\end{center}

 The following pair of complex-valued functions were introduced in \cite{lee} to describe the bubble
 break-off near the breaking point
\begin{equation}\label{car}
X(\lambda):=\dfrac{1}{2}\Big(\lambda+\mS(\lambda)\Big),\quad
Y(\lambda):=\dfrac{1}{2\,i}\Big(\lambda-\mS(\lambda)\Big).
\end{equation}
They  analytically extends the Cartesian coordinates $(X,Y)$ of the
interface $\gamma$
\begin{equation}\label{car1}
X=Re\, \lambda,\quad Y=Im\, \lambda,\quad \lambda\in\gamma,
\end{equation}
and allow to write the evolution law \eqref{dar} in the form
\begin{equation}\label{dar1}
\partial_t\, Y(X)=-\partial_X\,\Phi(X).
\end{equation}

The analysis of \cite{lee} concludes that after the break-off the
local structure of a small part of the interface containing the tips
of the bubbles falls into universal classes characterized by two
even integers $(4\,n, 2),\, n\geq 1,$ and a finite number $2n$ of
real deformation parameters $t_k$. By assuming symmetry of the curve
with respect to the $X$-axis, the general solution for the curve and
the potential in the $(4\,n, 2)$ class are
\begin{equation}\label{den}
Y:=\Big(\dfrac{U_X}{\sqrt{(X-a)(X-b)}}\Big)_\oplus\,\sqrt{(X-a)(Y-b)},\quad
 \Phi=-\sqrt{(X-a)(Y-b)},
\end{equation}
where $a$ and $b$ are the positions of the bubbles tips and
\begin{equation}\label{U}
U(X,\ct):=\sum_{j=1}^{2n} t_{j+1}\,X^{j+1}.
\end{equation}
Here the subscript $\oplus$ denotes the projection of $X$-series on
the positive powers. Due to the physical assumptions of the problem,
the function $Y$ inherates two conditions for its expansion  as
$X\rightarrow\infty$ \begin{equation}\label{hoh}
Y(X)=\sum_{j=1}^{2n}
(j+1)\,t_{j+1}\,X^{j}+\sum_{j=0}^{\infty}\dfrac{Y_n}{X^n}.
\end{equation}
which determine the positions $a$, $b$ of the tips. The conditions
are
\begin{enumerate}
\item From \eqref{dir} $\Phi\rightarrow -i\,\log \lambda$ as $\lambda \rightarrow \infty$. Hence \eqref{dar1} implies that the constant term $Y_0$ in \eqref{hoh} should be equal to $t$.
\item The coefficient $Y_1$ in front of $X^{-1}$ turns to be equal to $\Pi$, so that it  must vanish for a break-off \cite{lee}.
\end{enumerate}
As it was shown in \cite{lee}, imposing these two conditions on
\eqref{hoh} leads to a pair of  hodograph equations which arise in
the dispersionless AKNS hierachy. However, from \eqref{den} it is
straightforward to see that these equations coincide with the
hodograph equations \eqref{ho} associated with the system of string
equations \eqref{trii} provided one sets
\begin{align}\label{sett}
\nonumber X&=z,\quad Y=2\,m-V_z,\quad \Phi=z-u-2\,p,\\\\
\nonumber t_j&=0,\quad  \forall j\geq 2n+2;\quad t=t_1,\quad
x=\dfrac{\Pi}{2}=0.
\end{align}
For instance, we observe  that the evolution law  \eqref{dar1}
derives in a very natural form from the d1-Toda hierarchy. Indeed,
from \eqref{d5} and \eqref{den} we have
\[
p=\dfrac{1}{2}\,(z-u-\Phi),
\]
so that \eqref{sett} implies
\[
\partial_t Y=2\,\partial_{t_1}
m(z)-1=2\,\partial_{z}(z)_+-1
=2\,\partial_{z}(p+u)-1=-\partial_{z}\Phi=-\partial_{X}\Phi.
\]

In this way the integrable structure associated to the system of
string equations \eqref{trii} of the d2-Toda hierarchy manifests a
duality between the planar limit of the Hermitian matrix model and
the bubble break-off in Hele-Shaw cells. According to this
relationship the density of eigenvalues $\rho$ and the end-points
$a,\,b$ of its support in the Hermitian model are identified with
the interface function $Y$ and the positions of the bubbles tips,
respectively, in the Hele-Show model.

\vspace{0.5cm}

\noindent {\bf Acknowledgments}

\vspace{0.3cm}
The authors  wish to
thank the  Spanish Ministerio de Educacion y Ciencia (research project FIS2005-00319) and the European Science Foundation (MISGAM
programme) for their financial support.

\small


\label{lastpage}

\end{document}